# Intramolecular Structure of Proteins as driven by Steiner Optimization Problems

R.P.Mondaini, N.V.Oliveira – Alberto Luiz Coimbra Institute for Graduate Studies and Research in Engineering/COPPE – UFRJ – Technology Centre – 21.941-972 – P.O. Pox 68.511, Rio de Janeiro, RJ, Brazil.
Corresponding author: rpmondaini@gmail.com

*Abstract*

*In this work we intend to report on some results obtained by an analytical modelling of biomacromolecular structures as driven by the study of Steiner points and Steiner trees with an Euclidean definition of distance.*

## 1. Introduction

Ten years ago, the first trials of modeling the protein structure by a Steiner problem have appeared in the scientific literature [1]. The careful work with protein data banks led us to conclude that Nature had solved a NP-hard optimization problem by choosing a privileged Steiner tree topology. The Steiner Ratio Function (SRF) was introduced as an extension [2] of the Steiner Ratio after exhaustive computational experiments with a reduced search space version of Smith's algorithm for Euclidean Steiner trees [3].

## 2. The Steiner Ratio Function (SRF) Prescription for Helical Points sets

We now consider the set of discrete sets of points of a metric manifold whose points are evenly spaced along right circular helices. The subsequences formed by fixed external points and those formed by Steiner points, both of them obtained after skipping $(m-1)$ points are, respectively

$$(\mathrm{P}_j)_{m,\,l_{max}^j}:\quad P_j,\quad P_{j+m},\quad P_{j+2m},\quad ...,\quad P_{j+l_{max}^j\cdot m} \tag{1}$$

$$(\mathrm{S}_k)_{m,\,l_{max}^k}:\quad S_k,\quad S_{k+m},\quad S_{k+2m},\quad ...,\quad S_{k+l_{max}^k\cdot m} \tag{2}$$

where

$$l_{max}^j = [(n-j-1)/m];\; l_{max}^k = [(n-k-2)/m];\; 0 \le j \le m-1;\; 0 \le k \le m-1 \tag{3}$$

and the square brackets stand for the greatest integer value.
In the following we shall work in the 3-dimensional Euclidean space. The coordinates of the points above can be written

$$P_{j+lm} = (\cos(j+lm)\mathbf{w},\, \sin(j+lm)\mathbf{w},\, \mathbf{a}(j+lm)\mathbf{w});\quad 0 \le l \le l_{max}^j \tag{4}$$

$$S_{k+lm} = (r_m(\mathbf{w},\mathbf{a})\cos(k+lm)\mathbf{w},\, r_m(\mathbf{w},\mathbf{a})\sin(k+lm)\mathbf{w},\, \mathbf{a}(k+lm)\mathbf{w});\quad 0 \le l \le l_{max}^k \tag{5}$$

The helical point sets with $n$ and $n-2$ can be grouped into subsequences of the form (1) and (2) respectively. New sequences of $n$ and $n-2$ points can be defined by

$$\mathsf{P}_m = \bigcup_{j=0}^{m-1}(\mathrm{P}_j)_{m,\,l_{max}^j};\qquad \mathsf{S}_m = \bigcup_{k=1}^{m-1}(\mathrm{S}_k)_{m,\,l_{max}^k} \tag{6}$$

A 3-sausage's topology [2] is assumed to form Steiner trees (ST) with the points $P_{j+lm}$, $S_{k+lm}$, $\forall m$. The points $P_{j+lm}$ only are used to form spanning trees (SP). From the requirement of full Steiner trees, we can write

$$r_m(\mathbf{w},\mathbf{a}) = m\mathbf{a}\mathbf{w}/\sqrt{A_m(1+A_m)}\,;\;(1+A_m)^2 \geq m^2\mathbf{a}^2\mathbf{w}^2+1+A_m\,;\;A_m = 1-2\cos(m\mathbf{w}) \quad (7)$$

A straightforward calculation by using the first equation in (7) lead us to write the Euclidean lengths of these $m$-spanning trees and $m$-Steiner trees for $n \geq 1$, as

$$l_{SP}^m(\mathbf{w},\mathbf{a}) \approx n\sqrt{m^2\mathbf{a}^2\mathbf{w}^2+1+A_m}\,;\quad l_{ST}^m(\mathbf{w},\mathbf{a}) \approx n(1+m\mathbf{a}\mathbf{w}\sqrt{A_m/(1+A_m)}) \quad (8)$$

According to some previous results [3], the prescription for a Steiner Ratio Function should be written as

$$\mathbf{r}(\mathbf{a},\mathbf{w}) = \left(\min_m l_{ST}^m(\mathbf{a},\mathbf{w}) \Big/ \min_m l_{SP}^m(\mathbf{a},\mathbf{w})\right) \quad (9)$$

## 3. Restriction to Full Steiner Trees and the SRF Function

We can also see from the condition introduced by second equation in (7) that the surface for $m=1$ has the largest feasible domain. From the equations (7) and (9), we then have a proposal for the *SRF* function of a helical point set in the form

$$\mathbf{r}(\mathbf{a},\mathbf{w}) = \left(1+\mathbf{a}\mathbf{w}\sqrt{A_1/(1+A_1)}\right) \Big/ \min_m \sqrt{m^2\mathbf{a}^2\mathbf{w}^2+1+A_m} \quad (10)$$

If the Graham-Hwang's lowest bound is applied here or $\mathbf{r}(\mathbf{a},\mathbf{w}) \geq \sqrt{3}/3$, the corresponding $\mathbf{w}$-region is $\arccos(1/4) \leq \mathbf{w} \leq 2\mathbf{p} - \arccos(1/4)$. We can now present the optimization problem like

$$\mathbf{r}(\mathbf{a},\mathbf{w}) = \underset{m}{Max}\,\mathbf{r}_m(\mathbf{a},\mathbf{w}) = \underset{m}{Max}\left(\left(1+\mathbf{a}\mathbf{w}\sqrt{A_1/(1+A_1)}\right) \Big/ \sqrt{m^2\mathbf{a}^2\mathbf{w}^2+1+A_m}\right) \quad (11)$$

The first three surfaces of eq. (11) meet at possible global minimum of the surface given by eq. (10). The only non trivial solution is

$$\mathbf{w}_R = \mathbf{p} - \arccos(2/3)\,;\;\mathbf{a}_R = \sqrt{30}/[9(\mathbf{p}-\arccos(2/3))]\,;\;\mathbf{r}(\mathbf{w}_R,\mathbf{a}_R) = \frac{1}{10}(3\sqrt{3}+\sqrt{7}) \quad (12)$$

The last value corresponds to the value assumed by the authors of ref. [2] as their main conjecture, or $0.78419037337...$

## 4. The Weierstrass Theorem and the Existence of a Global Minimum

We define a compact domain by the region given below

$$\{H_{r_1} - \bigcup_{m\geq 2} H_{r_m}\} \cap \{(\mathbf{w},\mathbf{a})\,|\,\arccos(1/4) \leq \mathbf{w} \leq 2\mathbf{p}-\arccos(1/4)\} \quad (13)$$

where $H_{r_1}$, $H_{r_m}$ stands for the hypograph of the curves $r_1(\mathbf{w},\mathbf{a})=1$ and $r_m(\mathbf{w},\mathbf{a})=1$, $m \geq 2$, respectively. Since the surfaces $r_m(\mathbf{w},\mathbf{a})$, $m = 1, 2, 3$, are continuously decreasing from these boundaries of the region (13) towards their unique intersection point given by eq. (12), this point corresponds to a global minimum of surface (10). Q.E.D.